\documentclass[aps]{revtex4}

\usepackage{amssymb}
\usepackage[tbtags]{amsmath}
\usepackage{graphicx}
\usepackage{subfigure}

\DeclareMathOperator{\grad}{grad}
\newcommand{\abs}[1]{\lvert#1\rvert}

\begin{document}
\title{Phase Difference Function in Coherent Temporal-spatial Region\\
and Unified Equations of Steady, Non-steady Interference}

\author{Ji Luo}
\email{luoji@mail.ihep.ac.cn}
\affiliation{Accelerator Center, Institute of High Energy Physics, Beijing, China}

\date{\today}

\begin{abstract}
Phase difference function is established by means of phase transfer function between
time domains of source and interference point. The function reveals a necessary
interrelation between outcome of two-beam interference, source's frequency and measured
subject's kinematic information. As inference unified equations on steady and
non-steady interference are derived. Meanwhile relevant property and application are
discussed.
\end{abstract}

\maketitle

\section{Introduction}
On two beam interference, the explicit interrelationship among source's frequency
parameter, time functions of the beams between time domains of source and interference
point \cite{luoji:timefunction} and instantaneous outcome of the interference constructs a necessary
foundation for any two-beam interferometry's design and interpretation of measured
data. By means of phase transfer function between time domains of source and observer
\cite{luoji:transferrelation}, the accumulated phase or phase function for either of the wave beams at
any spatial point in the coherent temporal spatial region can be determined by source's
parameter and corresponding instant. Upon the beam's phase function in the coherent
temporal-spatial region, the phase difference function, two variables function in
interference temporal-spatial region, is established. Is the phase difference function
a general interrelationship itself. From it, the unified equations on steady and
non-steady interference are inferred directly under the cases respectively. For steady
interference, conventional rule on the interference spatial distribution is a
particular example of the unified equation as the two beam's wavelengths are same. In
addition Michelson-Morley experiment result is reinterpreted with the equation. For
non-steady interference two sets of the equations are derived for different
interferometry outcomes: beat frequency and fringe's instantaneous displacing velocity;
moreover on some of typical dynamical measurement: history of distance, velocity and
acceleration as well as source frequency property, the principle formulas are presented
for application illustration.

\section{Derivation of Phase Difference Function in Interferometry Temporal-spatial
Region}
In beam's transfer time-space $t' \in \tau_i'(r)$, $r \in V_i$, there is phase transfer functions (Ref.\ %
\cite{luoji:transferrelation}):
\begin{subequations}
\begin{equation}
\label{E:1a} \begin{split}
\varphi_i(r,t')&=\int_{t_{oi}'}^{t'}\omega_{i}'(r,\dot{t})\,d\dot{t}+\varphi_i(r,t_{oi}')\\
&=\int_{t_{oi}'-T_i(r,t_{oi}')}^{t'-T_i(r,t')}\omega_i(\dot{t})\,d\dot{t}+\varphi_i(t_{oi})\\
&=\varphi_i[t'-T_i(r,t')]=\varphi[t'-T_i(r,t')]
\end{split}
\end{equation}
here $
\varphi_i(r,t_{oi}')=\varphi_i[t_{oi}'-T_i(r,t_{oi}')]=\varphi_i(t_{oi})%
=\varphi[t_{oi}'-T_i(r,t_{oi}')] $, and
\begin{equation}
\label{E:1c}
\begin{split}\frac{\partial\varphi_i(r,t')}{\partial
t'}&=\omega_i'(r,t')=\omega[t'-T_i(r,t')]\frac{\partial t_i(r,t')}{\partial t'}\\
&=\omega[t'-T_i(r,t')]\left[1-\frac{\partial T_i(r,t')}{\partial t'}\right]
\end{split}
\end{equation}
\begin{widetext}
\begin{equation}
\label{E:1d}
\begin{split}\grad\varphi_i(r,t')&=-\omega[t'-T_i(r,t')]\grad T_i(r,t')\\
&=-\omega[t'-T_i(r,t')]\frac{1}{v_{ioi}(r,t')}\frac{\partial t_i(r,t')}{\partial
t'}\Vec n(\grad T_i)\\
&=-\omega[t'-T_i(r,t')]\frac{1}{v_{ioi}(r,t')}\left[1-\frac{\partial
t_i(r,t')}{\partial t'}\right]\Vec n(\grad T_i)
\end{split}
\end{equation}
\begin{equation}
\label{E:1e} \frac{\partial\varphi_i(r,t')}{\partial
t'}=\rho_{\varphi}(r,t')v_{ioi}(r,t')=\abs{\grad\varphi_i(r,t')}v_{ioi}(r,t')
\end{equation}
\end{widetext}
\end{subequations}

In coherent temporal-spatial region $r \in V_{c12} \in V_1 \cap V_2$, $t' \in
[\,t_0'(r), t_0'(r)+\tau_{c12}'(r)\,] \in \tau_1'(r)\cap \tau_2'(r)$; where $V_{c12}$ %
--- coherent spatial volume of wave beam 1 and 2, $\tau_{c12}'$ --- coherent time
interval of beam 1 and beam 2; two beams are split by same source but have different
main trajectories to reach interference point.

Now define the phase difference function in this coherent temporal-spatial region,
\begin{equation}\label{E:2}
\begin{split}\psi(r,t')&=\varphi_2(r,t')-\varphi_1(r,t')\\
&=\int_{t_0'}^{t'}\omega_2'(r,\dot{t})\,d\dot{t}+\varphi_2(r,t_0')-\int_{t_0'}^{t'}\omega_1'(r,\dot{t})\,d
\dot{t}-\varphi_1(r,t_0')\\
&=\varphi[t'-T_2(r,t')]-\varphi[t'-T_1(r,t')]\\
&=\int_{t'-T_1(r,t')}^{t'-T_2(r,t')}\omega(\dot{t})\,d\dot{t}
\end{split}
\end{equation}

\begin{widetext}
Thus there are the derivative property of the function:
\begin{equation}\label{E:3}
\begin{split}
\frac{\partial \psi(r,t')}{\partial t'}&=\omega_2'(r,t')-\omega_1'(r,t')\\
&=\omega[t'-T_1(r,t')]\frac{\partial T_1(r,t')}{\partial
t'}-\omega[t'-T_2(r,t')]\frac{\partial T_2(r,t')}{\partial
t'}+\omega[t'-T_2(r,t')]-\omega[t'-T_1(r,t')]
\end{split}
\end{equation}
\begin{equation}\label{E:4}
\begin{split}
\grad \psi(r,t')&=\grad \varphi[t'-T_2(r,t')]-\grad \varphi[t'-T_1(r,t')]\\
&=\grad \int_{t'-T_1(r,t')}^{t'-T_2(r,t')}\omega(\dot{t})\,d \dot {t}\\
&=\left\{-\omega[t'-T_2(r,t')]\frac{\partial T_2(r,t')}{\partial \Vec n (\grad
\psi)}+\omega[t'-T_1(r,t')]\frac{\partial T_1(r,t')}{\partial \Vec n (\grad
\psi)}\right\}\Vec n (\grad \psi)\\
&=\Biggl\{\biggl\{-\omega[t'-T_2(r,t')]\grad T_2(r,t')+\omega[t'-T_1(r,t')]\grad
T_1(r,t')\biggr\}\cdot\Vec n (\grad \psi)\Biggr\}\Vec n (\grad \psi)\\
&=\Biggl\{\biggl\{-\omega[t'-T_2(r,t')]\frac{\Vec n (\grad
T_2)}{v_{io2}(r,t')}\left[1-\frac{\partial T_2(r,t')}{\partial
t'}\right]\\
&\phantom{=\Bigg\{\bigg\{}+\omega[t'-T_1(r,t')]\frac{\Vec n (\grad
T_1)}{v_{io1}(r,t')}\left[1-\frac{\partial T_1(r,t')}{\partial t'}\right]\biggr\}\cdot
\Vec n (\grad \psi)\Biggr\}\Vec n (\grad \psi)
\end{split}
\end{equation}
\end{widetext}

where (See Appendix):
\[
\abs{\grad T(r,t')}=\frac{1}{v_{io}(r,t')}\frac{dt}{dt'}=\frac{1}{v_{io}(r,t')}\left[1-\frac{\partial
T(r,t')}{\partial t'}\right]
\]

\begin{equation}
\label{E:5}
\begin{split}
\frac{\partial \psi(r,t')}{\partial t'}&=\rho_{_{\!\mathrm{\psi}}}(r,t')v_{_{\mathrm{fringe}}}(r,t')\\
&=\abs{\grad \psi(r,t')}v_{_{\mathrm{fringe}}}(r,t')
\end{split}
\end{equation}

\section{Unified Equations of Steady and Non-steady Interference}
\renewcommand{\labelenumi}{\Alph{enumi}.}
\begin{enumerate}
\item When $\frac{\partial \psi(r,t')}{\partial t'}\equiv 0$ or $v_{_{\mathrm{
fringe}}}(r,t')\equiv 0$ and $\omega[t'-T_i(r,t')]=constant$, $\frac{\partial
T_i(r,t')}{\partial t'}=\frac{\partial T_i(r)}{\partial t'}=0$\\
there steady or static spatial distribution of interference
\begin{subequations}
\begin{equation}\label{E:6a}
\frac{2\pi}{\tau^*}[T_1(r)-T_2(r)]=\begin{cases}2k\pi&\quad\text{max.}\\
(2k+1)\pi&\quad\text{min.}\end{cases}\quad k=0,\pm 1,\dots
\end{equation}
or
\begin{equation}\label{E:6ab}
\int_r^{r+d}\abs{\grad\psi(r,t')}\,dr=2\pi \end{equation} \hfill $d$ -- short distance
between two fringes\hfill{}
\begin{equation}\label{E:6b}
T_1(r)-T_2(r)=\begin{cases}k\tau^*&\quad\text{max.}\\
(2k+1)\frac{\tau^*}{2}&\quad\text{min.}\end{cases}
\end{equation}

When $v_{io2}(r)=v_{io1}(r)=constant$, there is conventional rule on static spatial
distribution of steady interference
\begin{equation}\label{E:6c}
L_1(r)-L_2(r)=\begin{cases}k\lambda&\quad\text{max.}\\
(2k+1)\frac{\lambda}{2}&\quad\text{min.}\end{cases}
\end{equation}
\end{subequations}

\item For dynamical interferometry there exist general beat equation and fringe's
instantaneous displacement velocity equation as following
\begin{subequations}
\begin{equation}\label{E:7a}
\int_{t'}^{t'+\tau_b}\omega_b(r,\dot{t})\,d \dot
{t}=\int_{t'}^{t'+\tau_b}[\omega_2'(r,\dot{t})-\omega_1'(r,\dot t)]\, d\dot t=\pm 2\pi
\end{equation}
or
\begin{widetext}
\begin{equation}\label{E:7b} \int_{t'}^{t'+\tau_b}\left\{\omega[\dot t-T_2(r,\dot
t)]-\omega[\dot t-T_1(r,\dot t)]+\omega[\dot t-T_1(r,\dot t)]\frac{\partial T_1(r,\dot
t)}{\partial \dot t}-\omega[\dot t-T_2(r,\dot t)]\frac{\partial T_2(r,\dot t)}{\partial
\dot t}\right\}\, d \dot t=\pm 2\pi
\end{equation}
\end{widetext}
\begin{equation}\label{E:7c}
\int_{t'}^{t'+\tau_b}\left\{\omega[t_2(r,\dot t)]\frac{\partial t_2(r,\dot t)}{\partial
\dot t}-\omega[t_1(r,\dot t)]\frac{\partial t_1(r,\dot t)}{\partial \dot
t}\right\}\,d\dot t=\pm 2\pi
\end{equation}
\end{subequations}

\begin{equation}\label{E:8}
v_{_{\mathrm{fringe}}}(r,t')=-\frac{\partial \psi(r,t')}{\partial t'}\Vec n (\grad
\psi)\bigg/\abs{\grad \psi(r,t')}
\end{equation}

\item The principle formulas for a series of typically dynamical parameter measurement
\begin{figure}[htbp]
\subfigure[interferometer for measuring velocity and distance history]{\centering
\includegraphics[width=7cm]{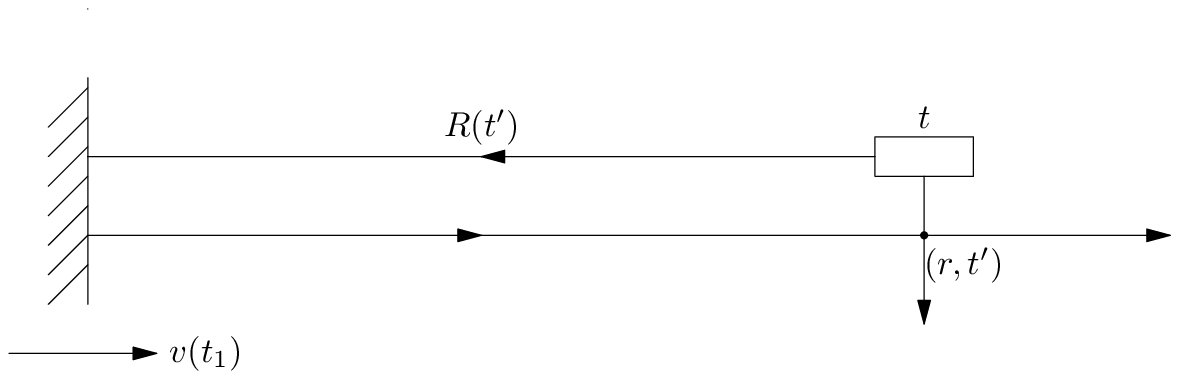}\label{F:1a:measure-v-r}}
\subfigure[interferometer for measuring acceleration history]{\centering
\includegraphics[width=7cm]{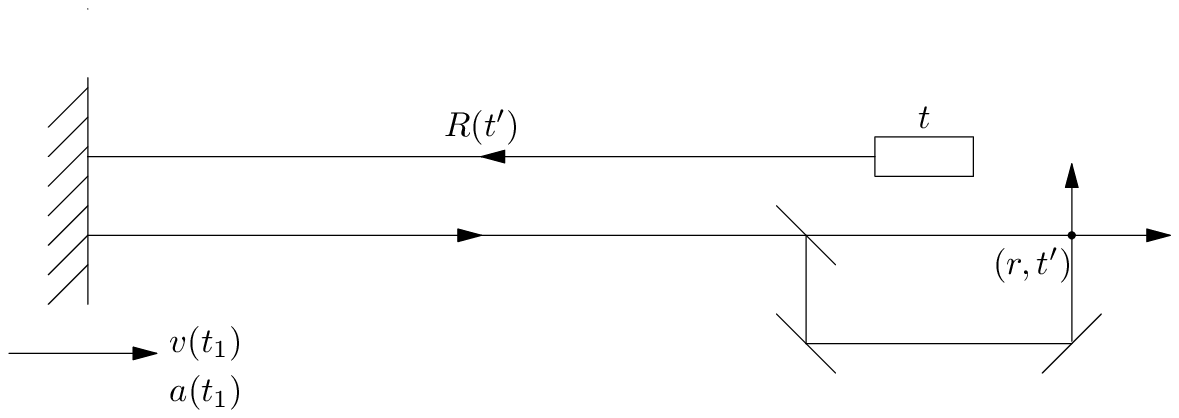}\label{F:1b:measure-a}}
\caption{Schematic of dynamic measurement}
\label{F:1:measurement}
\end{figure}

\renewcommand{\labelenumii}{(\roman{enumii})}
\begin{enumerate}
\item Ref.\ Fig.\ \ref{F:1a:measure-v-r}, $T_2(r,t')=T_2(r)$, $\omega(t)=\frac{2\pi}{\tau^*}=constant$,
from Eq.\ \eqref{E:7b}, there $\frac{2\pi}{\tau^*}\left.\frac{\partial
T_1(r,t')}{\partial t'}\right|_{\xi '}\tau_b=\pm 2\pi$ or $\left.\frac{\partial
T_1(r,t')}{\partial t'}\right|_{\xi '}=\frac{1}{\nu}\nu_b$, the $v(t_1)$ can be
resolved from $\frac{\partial T_1(t')}{\partial t'}$, here $T_1(t')=t'-t_1+t_1-t$.

\item To measure $R(t')$ history by two-beam interferometry with modulated frequency
source, Ref.\ Fig.\ \ref{F:1a:measure-v-r}, $\frac{d\omega(t)}{dt}=constant$, $T_2(t')=T_2\doteq 0$,
$T_1(t')-T_2(t')=\frac{2R(t')}{c}$, from Eq.\ \eqref{E:7c}
\begin{widetext}
\[
\int_{t'}^{t'+\tau_b}\left\{\int_{t(\dot t)-\frac{2R(\dot t)}{c}}^{t(\dot
t)}\frac{d\omega(\hat t)}{d\hat t}\,d\hat t+\omega\left[t(\dot t)-\frac{2R(\dot
t)}{c}\right]\frac{2}{c}\frac{dR(\dot t)}{d\dot t}\right\}\,d\dot t=2\pi
\]
\end{widetext}
\[
R(t')\doteq\left\{\pi
c\nu_b-\omega\left[t'-\frac{2R(t')}{c}\right]\frac{dR(t')}{dt'}\right\}\bigg/\frac{d\omega(t)}{dt}
\]

\item To measure $a(t_1)$ history by VISAR~\cite{goosman:laserinterfer,barker:laserinterfer,luoji:visar}, %
Ref.\ Fig.\ \ref{F:1b:measure-a},
$\frac{dT_1(t')}{dt'}=\left.\frac{dT_2(\dot t)}{d\dot t}\right|_{t'-\tau}$ and
$\omega(t)=\frac{2\pi}{\tau^*}=constant$;
from Eq.\ \eqref{E:7b} there $\omega \int_{t'}^{t'+\tau_b}\left[\frac{dT_2(\dot
t)}{d\dot t}-\left.\frac{dT_2(\dot t)}{d \dot t}\right|_{\dot t-\tau}\right]\,d\dot
t=2\pi$, or $ \left.\frac{d^2T_2(\dot t)}{d\dot t^2}\right|_{\xi '}=\frac{\nu_b}{\tau
\nu}$, $\xi '\in(t',t'+\tau_b)$.

The $a(t_1)$ can be resolved form $\left.\frac{d^2 T_2(\dot t)}{d\dot t^2}\right|_{\xi
'}$~\cite{luoji:visar}. Moreover from Eq.\ \eqref{E:8} there $v_{_{\mathrm{
fringe}}}(r,t')=2\pi\nu\tau\frac{d^2T_2(t')}{dt'^2}\big/\abs{\grad\psi}$, so $a(t_1)$ is
also resolved by fringe's displacing speed.

\item To measure the rate of radiant frequency of electron in magnetic dipole ---
$\frac{d\nu(t)}{dt}$, Ref.\ Fig.\ 2, insert high dispersion medium with length between
radiant source and interferometer. If $\frac{d\nu(t)}{dt}$ does not exist, short
distance of fringes will be independent to the medium length $L$; if
$\frac{d\nu(t)}{dt}$ exists, the fringes' distance will be proportional to the length
and refractive index of dispersion medium.
\end{enumerate}
\end{enumerate}

\section{Discussion}
\renewcommand{\labelenumi}{\Alph{enumi}.}
\begin{enumerate}
\item Phase difference function, through phase transfer function and involved time
function, reflects the theoretical interrelation between outcome of two-beam
interference, frequency characteristics of two autonomic wave sources, for example,
as beat equation
$\int_{t'}^{t'+\tau_b}\!\left\{\omega_2[t_2(r,\dot t)]\frac{\partial t_2(r,\dot t)}{\partial
\dot t}-\omega_1[t_1(r,\dot t)]\frac{\partial t_1(r,\dot t)}{\partial \dot
t}\right\}\,d\dot t=\pm 2\pi$
and subject's
kinematic information involved in corresponding time function. The time function
applied in phase transfer function must be reversible, in most cases it is positive
reversible, that is, $\frac{dt'}{dt}=1+\frac{dT(t')}{dt}>0$; all inference and property
of the phase difference function can be applied in case of two autonomic wave sources,
although in most cases two coherent wave sources are derived from one by splitting.
Since phase difference function is consisted of or can be expressed by spatial frame
independent physical quantities, the theoretical relation and consequent inferences do
no depend upon spatial-frame's selection.

\item Explanation of zero fringe displacing result of Michelson-Morley experiment.

According to steady interference equation \eqref{E:6a} \eqref{E:6ab}, the only existing
explanation for the experimental result is phase difference at observing interference
point remains same or invariable in two cases, that is, light speed with respect to
interferometer remains same in both cases.
\end{enumerate}

\section{Conclusion}
Phase difference function reveals the necessary interrelation between outcome of
two-beam interference, frequency parameter of two autonomic wave sources, and concrete
time function's information affected by subject kinematic movement. Unified equation on
steady and non-steady two-beam interference can be derived from the phase difference
function. Phase difference function and related inference are independent to
spatial-frame's selection or remain invariable under frame transformation \cite{luoji:frametransform}.

\appendix*
\section{Derivation of $\abs{\grad T(r,t')}$}
\label{appendix1}
\[
\begin{split}
\abs{\grad T(r,t')}&=\lim_{\Delta r \to 0}\frac{T(r+\Delta r,t')-T(r,t')}{\Delta
r}\\
&=\lim_{\Delta r \to 0}\frac{t'-T(r,t')-[t'-T(r+\Delta r,t')]}{\Delta r}\\
&=\lim_{\Delta r \to 0}\frac{t(r,t')-t(r+\Delta r,t')}{\Delta r}=\lim_{\Delta r \to
0}\frac{1}{\frac{\Delta r}{\Delta t'}}\frac{\Delta t}{\Delta t'}\\
&=\frac{1}{v_{io}(r,t')}\frac{dt}{dt'}=\frac{1}{v_{io}(r,t')}\left[1-\frac{\partial
T(r,t')}{\partial t'}\right]
\end{split}
\]
or from
\[
\begin{split}
\abs{\grad T(r,t')}&=\abs{\grad T[r,t'-T(r,t')]}=\abs{\grad T[r,t(r,t')]}\\
&=\frac{\partial T(r,t')}{\partial r}=\frac{d T[r,t(r,t')]}{dr}
\end{split}
\]
and
\[
\begin{split}
\frac{\partial T(r,t)}{\partial r}&=\lim_{\Delta r \to 0}\frac{T(r+\Delta
r,t)-T(r,t)}{\Delta r}\\
&=\lim_{\Delta r \to 0}\frac{t+T(r+\Delta r,t)-[t+T(r,t)]}{\Delta
r}\\&=\lim_{\Delta r \to 0}\frac{\Delta t'}{\Delta r}=\frac{1}{v_io(r,t)}
\end{split}
\]
there (Ref.\ \cite{luoji:timefunction})
\[
\begin{split}
\abs{\grad T(r,t')}&=\frac{\partial T(r,t')}{\partial r}=\frac{\partial
T(r,t)}{\partial r}\Bigg/\left[1+\frac{\partial T(r,t)}{\partial t}\right]\\
&=\frac{1}{v_{io}(r,t)}\left[1-\frac{\partial T(r,t')}{\partial t'}\right]
\end{split}
\]

\end{document}